\documentclass{aastex631}

\usepackage[utf8]{inputenc}
\usepackage{amsmath, amssymb, amsthm}
\usepackage{geometry}
\usepackage{hyperref}
\usepackage{physics} 
\usepackage{booktabs} 
\usepackage[rightcaption]{sidecap}   
\usepackage{color} 

\accepted{\today}

\shorttitle{Inner 3D slope from projected surface density}
\shortauthors{S\'anchez Almeida}
\graphicspath{{./}{figures/}}

\begin{document}

\title{Accurate inner stellar density slopes from projected surface densities in galaxies}

\correspondingauthor{J. S\'anchez Almeida} 
\email{jos@iac.es}

\author[0000-0003-1123-6003]{Jorge S\'anchez Almeida}  \affil{Instituto de Astrof\'\i sica de Canarias, La Laguna, Tenerife, E-38200, Spain} \affil{Departamento de Astrof\'\i sica, Universidad de La Laguna, Tenerife, Spain}




\begin{abstract}
  The inner slope of the three-dimensional stellar density in dwarf galaxies ($\rho'[0]$) is a sensitive probe of possible departures from the collisionless cold dark matter (CDM) paradigm, since cored stellar distributions ($\rho'[0]=0$) cannot easily reside within the cuspy potentials CDM predicts for low-mass systems. Photometry alone offers an observationally inexpensive way to constrain $\rho'(0)$, making this approach particularly attractive for the faint galaxies most relevant to dark matter (DM) studies. Inferring volume densities, however, requires deprojecting the observed stellar surface density, $\Sigma(R)$, a procedure that is notoriously ambiguous in the presence of noise. To avoid explicit deprojection, we derive an expression (Eq.~[\ref{eq:finally2_old}]) to obtain $\rho'(0)$ directly from the radial derivatives of $\Sigma(R)$, assuming spherical symmetry and smooth finite density profiles. All projected profiles are shown to have the same central functional form, independent of the underlying volume density (Eq.~[\ref{eq:sigma11}]). As a result, the derivatives of $\Sigma(R)$ can be extrapolated to the center using constraints from larger radii, which in turn yields $\rho'(0)$.  As an illustration, we apply the method to six ultra-faint dwarf (UFD) galaxies, finding that all of them have a surface density with the same shape, from which the presence of stellar cores is inferred ($\rho'[0]\simeq 0$). The technique also has the ability to diagnose $\rho'[0]>0$, corresponding to galaxies with a central stellar mass deficit potentially linked to black-hole scouring, MONDian dynamics, or deviations from CDM.
\end{abstract}

\keywords{
Cold dark matter (265) ---
Dark matter (353) ---
Dark matter distribution (356) ---
Dwarf galaxies (416) ---
Star counts (1568)  
}


\section{Introduction} \label{sec:intro}

The inner slope of the three-dimensional stellar distribution in dwarf galaxies is emerging as a fundamental quantity for assessing deviations of the dark matter (DM) from the collisionless cold dark matter (CDM) paradigm. It plays a key role on the latest reincarnation of the classical core–cusp problem \citep[e.g.,][]{2017ARA&A..55..343B,2021Galax...9..123D}. If the stellar volume density $\rho$ of a galaxy has a core, namely, if
  \begin{equation}
   \rho'(r) = \frac{d\rho(r)}{d r}\simeq 0 ~{\rm when}~r\to 0,
    \label{eq:definition}
\end{equation}
then the DM potential hosting the stars can hardly has the inner cusp predicted by CDM \citep[a NFW potential, as described  by][]{1997ApJ...490..493N}.
The inconsistency arises because this combination of stars and DM requires an unphysical distribution function that becomes negative somewhere in the phase space. The non-negativity of the distribution function is a long-standing consistency requirement in models of gravitating stellar systems \citep[e.g.,][]{1962MNRAS.123..447L,1990ApJ...356..359H,1992MNRAS.255..561C,2006ApJ...642..752A}. From these early works, it is known that two-component models combining quasi-isothermal density profiles (cores) and \citeauthor{1948AnAp...11..247D} profiles (more cuspy) face consistency problems, as they produce negative distribution functions near the center \citep{1992MNRAS.255..561C,1999ApJ...520..574C}.
\citet{2006ApJ...642..752A} established a theorem, later generalized by \citet{2010MNRAS.408.1070C}, relating the central logarithmic slope of the density profile to the anisotropy of the stellar velocity distribution. In particular, it rules out core density profiles embedded in NFW potentials for isotropic stellar orbits.
\citet{2023ApJ...954..153S} proposed exploiting this tension in the context of DM nature studies, as it allows the presence of NFW halos to be discarded from photometry alone, which is particularly suitable for objects too faint to be accessible by standard techniques.
Strictly speaking, this mathematical incompatibility discussed by \citet{2023ApJ...954..153S}
holds for spherically symmetric systems with stars in isotropic orbits, but these assumptions can be relaxed to axi-symmetric systems \citep{2024A&A...690A.151S}, radially biased orbits \citep{2023ApJ...954..153S}, and even potentials with small cores \citep{1992MNRAS.255..561C,2024RNAAS...8..167S}. This incompatibility is important because the CDM cosmology predicts the DM halos of small galaxies ($M_\star< 10^6 M_\odot$) to be cuspy independently of the stellar feedback \citep[e.g.,][]{2012ApJ...759L..42P,2016MNRAS.456.3542T} so, in principle, finding stellar cores in such galaxies would imply the need to go beyond CDM, as it has been suggested already \citep{2024ApJ...973L..15S}. The strength of this approach for diagnosing whether the DM is CDM or not lies in its reliance on photometry alone, which is particularly valuable for faint galaxies below the critical mass threshold, where traditional spectroscopic techniques for recovering the shape of dark matter halos are challenging. Therefore, this method has the potential to increase the available statistics with data from ongoing deep imaging surveys like those provided by the Euclid satellite \citep{2011arXiv1110.3193L} or the Rubin Observatory \citep{2019ApJ...873..111I}.  

In essence, the technique needs to recover the central density gradient (Eq.~[\ref{eq:definition}]) from the measured projected stellar surface density $\Sigma(R)$. Assuming spherically symmetry, $\Sigma(R)$ is just the Abel transform of $\rho(r)$,  
\begin{equation}
  \Sigma (R) = 2 \int_R^\infty \frac{\rho(r)\,r}{\sqrt{r^2 - R^2}}dr,
  \label{eq:abel}
  \end{equation}
  with $R$ the projected distance from the center. Thus, to recover $\rho'(0)$ from the observed $\Sigma(R)$ one has to apply the inverse Abel transform,
\begin{equation}
  \rho(r)=-\frac{1}{\pi}\int_r^\infty \frac{\Sigma'(R)}{\sqrt{R^2-r^2}}dR.
  \label{eq:abel_inv}
\end{equation}
The direct numerical usage of Eq.~(\ref{eq:abel_inv}) to recover $\rho'(0)$ is prone to systematic errors for a number of reasons, among others:
(a) catastrophic loss of precision when evaluating the integrand close to the lower limit, which contains a singularity,
(b)  numerical noise dominates derivatives of $\Sigma$ at small radii,
(c) differentiation amplifies the errors in $\Sigma$,
(d) since the transform is defined until $\infty$, extrapolation at large radii affects the result, and truncation of the integral introduces systematic bias,
(e) $\Sigma$ is known only at discrete points so that interpolation is required which introduces kinks and oscillations in the result,
and, thus, (f)  error estimates are hard to quantify rigorously.
The present work shows a procedure to overcome these difficulties using an approximate analytic solution to the Abel inversion problem, that holds where it is needed at $r\to 0$. We also show how it can be applied to observed surface densities to infer the inner slope of three-dimensional distributions of stars.
In addition to spherical symmetry, this procedure assumes the volume density $\rho(r)$ to be finite, of finite derivative, and of finite extent. While these assumptions are restrictive, they are expected to represent real stellar systems, where infinites are physically unlikely and spherical symmetry is often a reasonable approximation.

The paper is organized as follows:
Sect.~\ref{sec:central_volume} works out the expression for $\rho'(r)$ when $r\to 0$ as a function of the potentially observable variables $\Sigma'(R)$ and $\Sigma''(R)$. These observables are prone to severe noise so, in Sect.~\ref{sec:shapes},  we work out the general expression for the dependence of $\Sigma(R)$ on $R$ when $R\to 0$. It happens to be independent of the function $\rho(r)$, therefore,  it can be used to extrapolate the observed $\Sigma(R)$ to $R\to 0$, allowing us to derive $\Sigma'(0)$ and $\Sigma''(0)$ and so $\rho'(0)$. To illustrate the feasibility of this approach, Sec.~\ref{sec:ufds} applies it to infer $\rho'(0)$ for the Ultra Faint Dwarfs (UFDs) observed by \citet{2024ApJ...967...72R} and  analyzed in \citet{2024ApJ...973L..15S}. Section~\ref{sec:discussion} summarizes the results and discusses the potential of the technique.

%
\section{Derivative of the central volume density from surface densities}\label{sec:central_volume}

Changing variables to $u^2=R^2-r^2$, the inverse Abel transform in Eq.~(\ref{eq:abel_inv}) becomes
\begin{equation}
  \rho(r) = -\frac{1}{\pi}\int_{0}^\infty\frac{\Sigma'(\sqrt{u^2+r^2})}{\sqrt{u^2+r^2}}du,
\end{equation}
where the integration limits are constant and derivatives can be taken directly. Using this expression, 
\begin{equation}
  \rho'(r) =-\frac{r}{\pi} \Bigg[
    \int_0^\infty \frac{\Sigma''(\sqrt{u^2+r^2})}{u^2+r^2}du-
    \int_0^\infty \frac{\Sigma'(\sqrt{u^2+r^2})}{(u^2+r^2)^{3/2}} du
  \Bigg].
\label{eq:finally1}
\end{equation}
Assuming $\Sigma''(r)\not=0$ when $r\to 0$, then
\begin{equation}
  r\,\int_0^\infty \frac{\Sigma''(\sqrt{u^2+r^2})}{u^2+r^2}du\simeq r\,\Sigma''(r) \int_0^{r_0} \frac{1}{u^2+r^2}du+ r\,\int_{r_0}^\infty \frac{\Sigma''(u)}{u^2}du \simeq \Sigma''(r)\atan(r_0/r),
  \label{eq:uno}
\end{equation}
where $r_0$ has been chosen small enough for $\Sigma''(\sqrt{u^2+r^2})$ to be $\sim \Sigma''(r) $ in the first integral whereas the dependence on $r$ of the second integral is neglected since the integrand never diverges and $r\to 0$.  Similarly,
\begin{equation}
  r\, \int_0^\infty \frac{\Sigma'(\sqrt{u^2+r^2})}{(u^2+r^2)^{3/2}}du\simeq   \Sigma'(r) \int_0^{r_0} \frac{1}{u^2+r^2}du+ r\,\int_{r_0}^\infty\frac{\Sigma'(u)}{u^3}du \simeq
  \frac{\Sigma'(r)}{r}\atan(r_0/r),
  \label{eq:dos}
\end{equation}
where we have assumed that $\Sigma(\sqrt{u^2+r^2})/\sqrt{u^2+r^2}\simeq \Sigma'(r)/r\not=0$ when $r\to 0$ in  the first integrand. Note that $\Sigma'(r)$ could not be pullet out of the first integral in Eq.~(\ref{eq:dos}) since it is not constant when $r\to 0$, whereas  $\Sigma'(r)/r$ is (to be shown in Sect.~\ref{sec:shapes}). Using the approximations in Eqs.~(\ref{eq:uno}) and (\ref{eq:dos}), Eq.~(\ref{eq:finally1}) reduces to
  \begin{equation}
    \rho'(r)\simeq \frac{\atan(r_0/r)}{\pi}\left[\Sigma'(r)/r- \Sigma''(r)\right],
      \label{eq:finally2}
    \end{equation}
and noting that $\atan(r_0/r)\to \pi/2$ as $r\to 0$, it becomes\footnote{The referee pointed out a more direct derivation of Eq.~(\ref{eq:finally2_old}),  integrating Eq.~(\ref{eq:abel_inv}) by parts with respect to $d\sqrt{R^2 - r^2}$, and then carrying out a standard asymptotic expansion for $r\to 0$.}
\begin{equation}
    \rho'(r) \simeq \frac{1}{2}\left[\Sigma'(r)/r- \Sigma''(r)\right],
      \label{eq:finally2_old}
    \end{equation}
which allows us to infer $\rho'(0)$ from the observed $\Sigma'(R)/R$ and $\Sigma''(R)$ when $R\to 0$.

\begin{figure*}[ht!] 
\centering
\includegraphics[width=0.65\linewidth]{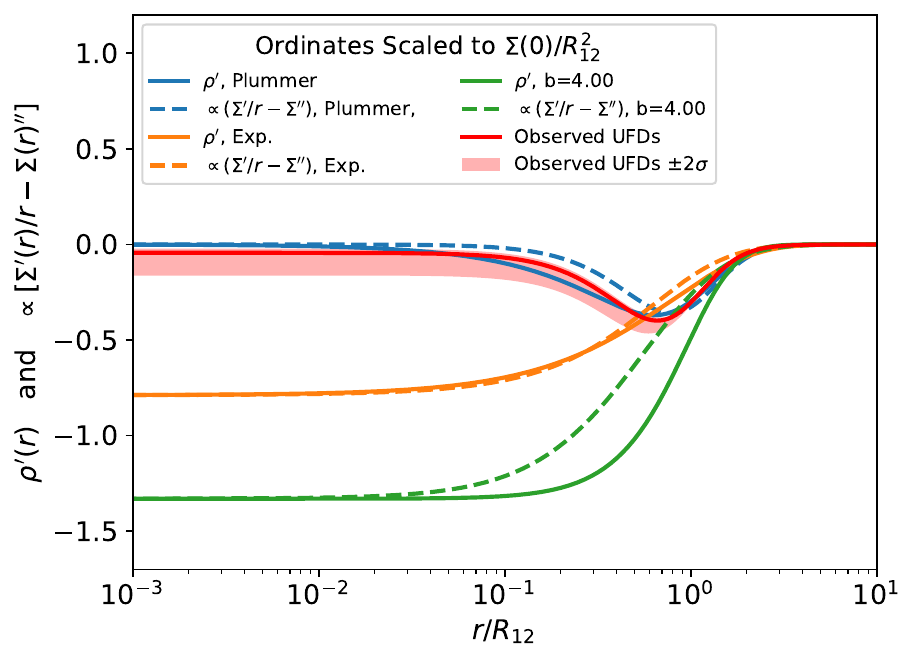}
\caption{
  Testing Eq.~(\ref{eq:finally2}) including density profiles with $\rho'(0)=0$ (Schuster-Plummer, Eq.~[\ref{eq:plummer}]) and with $\rho'(0)\not=0$ (exponential, Eq.~[\ref{eq:exp}], and shifted power law, Eq.~[\ref{eq:shift_power}]). The solid lines are the true $\rho'(r)$ whereas the dashed lines correspond to $\atan(r_0/R)\times[\Sigma'(R)/R-\Sigma''(R)]/\pi$.
We use $r_0=r_s$ for plotting, but this parameter has little influence on the result. The red solid line and shaded region show the range of values inferred from the UFDs in \citet{2024ApJ...967...72R},  as explained in Sect.~\ref{sec:ufds}.
}
\label{fig:central_dip9_pub}
\end{figure*}
To check the goodness of the approximation in Eqs.~(\ref{eq:finally2}) and (\ref{eq:finally2_old}), Fig.~\ref{fig:central_dip9_pub} shows the two sides of Eq.~(\ref{eq:finally2}) for density profiles having $\rho'(0)\not= 0$ and $\rho'(0) = 0$ -- the solid and the dashed lines correspond to $\rho'$ and $\atan(r_0/r)\,[\Sigma'/r-\Sigma'']/\pi$, respectively. Figure~\ref{fig:central_dip9_pub} includes a Schuster-Plummer profile ($\rho'[0]= 0$; the blue lines),
\begin{equation}
\rho(r)=\frac{\rho_s}{\left[1+(r/r_s)^2\right]^{5/2}},
\label{eq:plummer}
\end{equation}
an exponential profile ($\rho'[0]\not= 0$; the orange lines),
\begin{equation}
\rho(r) = \rho_s \exp\left[-(r/r_s)\right],
\label{eq:exp}
\end{equation}
and a displaced power law profile (the green lines),
\begin{equation}
 \rho(r)=\frac{\rho_s}{(1+r/r_s)^4},
\label{eq:shift_power}
\end{equation}
which also has $\rho'(0)\not= 0$ but larger than the value in an exponential. The symbols $\rho_s$ and $r_s$ are constants parameterizing the central density and the width of the distribution, respectively. In all three cases, the central surface density scales as $\rho_s r_s$ whereas the half-central density radius $R_{12}$, defined as
\begin{equation}
  \Sigma(R_{12})=\frac{1}{2}\Sigma(0),
  \label{eq:R12}
\end{equation}
scales as $r_s$. Thus, the scaling to $\Sigma(0)/R_{12}^2$ (ordinates) and $R_{12}$ (abscissas) used in Fig.~\ref{fig:central_dip9_pub} makes the plot independent of $r_s$ and $\rho_s$. Moreover, the parameters used in such scaling can be readily inferred provided $\Sigma(R)$ is observed. The functional form in Eqs.~(\ref{eq:plummer}) and (\ref{eq:exp}) has the advantage that the associated $\Sigma(R)$ and its derivatives admit closed-form expressions (Appendix~\ref{sec:appa}), enabling a thorough validation of our numerical algorithms to compute $\Sigma'(R)$ and $\Sigma''(R)$.

Figure~\ref{fig:central_dip9_pub} shows that Eq.~(\ref{eq:finally2}) is accurate enough to discriminate between profiles with cores (blue line) and without cores (orange and green lines) for $r \lesssim 0.3\,R_{12}$. At radii smaller than this limit, the differences between $(\Sigma'/R-\Sigma'')/2$ and $\rho'$ are smaller than the differences between the profiles with and without cores. One may think that these radii are too small to allow a proper determination from the observed $\Sigma(R)$. However, as we show in Sect.~\ref{sec:shapes}, the innermost regions of $\Sigma(R)$ have a well defined dependence on $R$ independently of $\rho(r)$, which allows us to extrapolate $\Sigma(R)$ from outer radii all the way down to $R=0$, thus providing $\rho'(0)$ through Eq.~(\ref{eq:finally2}) from real observed surface density profiles.  

%
\section{General form of the surface density profiles at small radii}\label{sec:shapes}

In order to derive the shape of the surface density profile in the innermost regions, we repeat the analysis carried out in the previous section but this time to infer $\Sigma(R)$ by integration of $\Sigma'(R)$. In order to use the Abel transform to compute $\Sigma'(R)$, it is convenient to transform Eq.~(\ref{eq:abel}) into an integral with constant limits.  The change of variables $l^2=r^2-R^2$ renders 
\begin{equation}
  \Sigma (R) = 2 \int_0^\infty \rho(\sqrt{l^2+R^2}) \, dl,
  \label{eq:def1}
  \end{equation}
so that
\begin{equation}
\Sigma'(R) = 2R\, \int_0^\infty \frac{\rho'(\sqrt{l^2+R^2})}{\sqrt{l^2+R^2}}\, dl.
    \label{eq:deriv1}
  \end{equation}
Following the thread of arguments in Sect.~\ref{sec:central_volume},  the integral in the right-hand-side of Eq.~(\ref{eq:deriv1}) can be approximated for $R\to 0$, provided $\rho$ is a well behaved function with finite $\rho'(0)$. When $R\to 0$
\begin{equation}
 \int_0^\infty \frac{\rho'(\sqrt{l^2+R^2})}{\sqrt{l^2+R^2}}\, dl \simeq
\rho'(R)  \int_0^{R_0} \frac{dl}{\sqrt{l^2+R^2}} +
\int_{R_0}^\infty \frac{\rho'(\sqrt{l^2+R^2})}{\sqrt{l^2+R^2}}\, dl,
\label{eq:split}
\end{equation}
with $R_0$ an arbitrary distance as small as needed for $\rho'(l)$ to be replaced with $\rho'(R)$ in the first integral.  The first integral can be integrated analytically to yield,
\begin{equation}
  \int_0^{R_0} \frac{dl}{\sqrt{l^2+R^2}} = 
  \ln(R_0+\sqrt{R_0^2+R^2})-\ln R \simeq \ln(2R_0/R).
     \label{eq:int1_good}
\end{equation}
The second integral in Eq.~(\ref{eq:split}) is finite and independent of $R$ when $R\to 0$,
\begin{equation}  
\int_{R_0}^\infty \frac{\rho'(\sqrt{l^2+R^2})}{\sqrt{l^2+R^2}}\, dl \simeq \int_{R_0}^\infty \frac{\rho'(l)}{l}\, dl \equiv I'(R_0).
       \label{eq:the_tail}
\end{equation}
Using Eqs.~(\ref{eq:deriv1}), (\ref{eq:int1_good}), and (\ref{eq:the_tail}), and maintaining the leading terms in $R\ln R$ and $R$, one finds
\begin{equation}
  \Sigma'(R)\simeq 2\rho'(0)\,R\ln(2R_0/R)+2R I'(R_0),
  \label{eq:first_derivative}
\end{equation}
which can be integrated analytically
to yield
\begin{equation}
  \Sigma(R) = \Sigma(0)+\Sigma_1\,R^2\ln R+ \Sigma_2\,R^2+\dots. 
  \label{eq:sigma11}
\end{equation}
The dots at the end of the previous expression represent terms that go to zero faster than $R^2$ when $R\to 0$, and
\begin{equation}
  \Sigma_1 = -\rho'(0),
  \label{eq:sigma11a}
\end{equation}
with
\begin{equation}
  \Sigma_2 = \rho'(0)\,\left[\ln(2R_0)+\frac{1}{2}\right]+I'(R_0).
  \label{eq:sigma11b}
\end{equation}
Note that the expression for $\Sigma(R)$ in Eq.~(\ref{eq:sigma11}) satisfies
\begin{equation}
  \frac{1}{2}\left[\Sigma'(R)/R-\Sigma''(R)\right] = -\Sigma_1+ \dots =\rho'(0)+\dots,
  \label{eq:referee1}
\end{equation}
which corresponds to the first order term when expanding $\rho'(R)$ in
Eq.~(\ref{eq:finally2_old}). As shown in Appendix~\ref{sec:appc}, including the $R^3$ term in Eq.~(\ref{eq:sigma11}) introduces a linear contribution, $\rho''(0)\,R$, on the right-hand side of Eq.~(\ref{eq:referee1}).

\begin{figure*}[ht!] 
\centering
\includegraphics[width=0.7\linewidth]{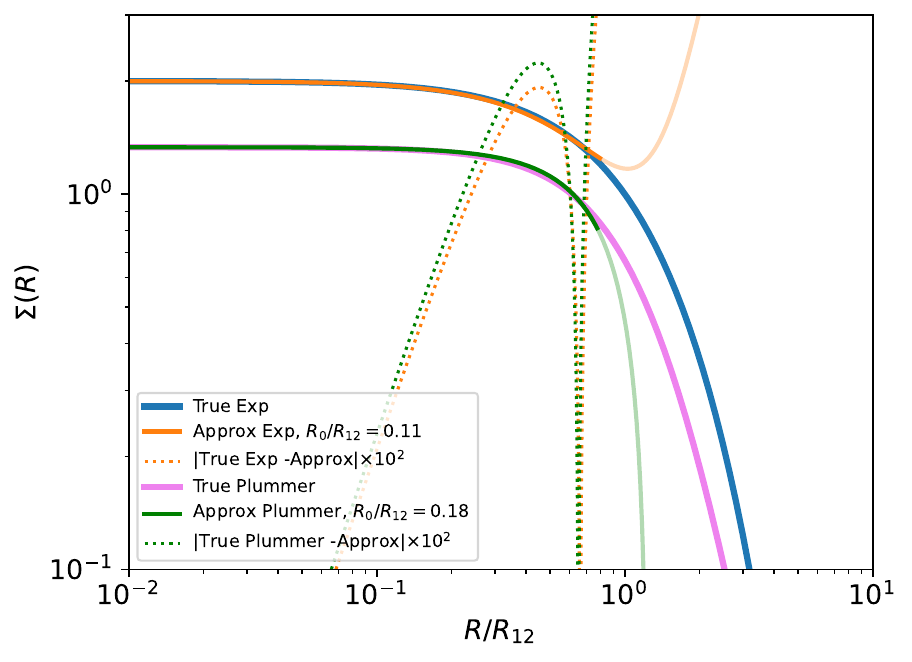}
\caption{
  Intended to illustrate the approximation contained in Eqs.~(\ref{eq:sigma11}),  (\ref{eq:sigma11a}), and (\ref{eq:sigma11b}). The blue solid line is the projected surface density of an exponential profile (Eq.~[\ref{eq:exp}]). The orange solid line is the approximation, derived from a fit of the true profile in the central region leaving $R_0$ as free parameter  -- see its value in the inset.  The orange line becomes more transparent beyond the radius used for fitting. The residuals, magnified by a factor of one hundred, are shown as the orange dotted line. The same code is used to represent the surface density corresponding to a Schuster-Plummer profile (Eq.~[\ref{eq:plummer}]): the violet solid line is the true surface density, the green solid line is the approximation, and the green dotted line represents the residuals scaled up. 
All radii are normalized to the radius at half the central surface density (Eq.~[\ref{eq:R12}]).
}
\label{fig:central_dip8}
\end{figure*}
Figure~\ref{fig:central_dip8} illustrates the approximation worked out above in the case of a Schuster-Plummer profile (Eq.~[\ref{eq:plummer}]) and an exponential (Eq.~[\ref{eq:exp}]). We fit the central region of the corresponding surface density using Eqs.~(\ref{eq:sigma11}), (\ref{eq:sigma11a}), and (\ref{eq:sigma11b}) using $R_0$ as free parameter. True functions, approximations, and residuals are included in the figure, as indicated in the inset. The approximation is within one percent of the true surface density for $R\lesssim R_{12}$. Even if this is not shown in the figure, the approximations are  similar when  $\Sigma_1$ and $\Sigma_2$ are treated as independent free parameters to fit Eq.~(\ref{eq:sigma11}).

There are several conclusions to be drawn from Eq.~(\ref{eq:first_derivative}):
(a) Independently of the exact value of $\rho'(0)$,  $\Sigma'(R)\to 0$ when $R\to 0$. Thus all well behaved volume density profiles produce surface densities with cores (i.e., with $\Sigma'[0]\to 0$), as shown in fairly general terms by \citet{2025arXiv251215886V}. The difference between surface densities with and without three-dimensional cores lies in the relative weight of the two last terms in Eq.~(\ref{eq:sigma11}). In the limit where the volume density has a core (i.e., $\rho'[0]=0$), the dependence of $\Sigma$ on $R$ is purely quadratic.
(b) Equation~(\ref{eq:first_derivative}) implies that all surface density profiles from spherically symmetric finite volume densities have the same shape at the center (Eq.~[\ref{eq:sigma11}]). This suggests that $\rho'(0)$ can be inferred using Eq.~(\ref{eq:finally2}) by extrapolating to the center the function $\Sigma(R)$ observed off the center. This procedure is illustrated in Sect.~\ref{sec:ufds} with observations of UFD galaxies.
(c) In addition to $\rho'(0)$, the surface density depends on $I'(R_0)$, which integrates the whole density profile to infinity (Eq.~[\ref{eq:the_tail}]). Thus, even when $R$ is small, the coefficients defining $\Sigma(R)$ depend on $\rho(r)$ at all radii.
(d) Equations~(\ref{eq:first_derivative}) and (\ref{eq:sigma11}) hold even when $\rho'(0) > 0$, which is a possibility discussed in some detail in Sect.~\ref{sec:discussion}.

%

\section{Application to UFD galaxies}\label{sec:ufds}
\citet{2024ApJ...973L..15S} showed that six UFD galaxies follow a stellar surface density profile that, after trivial scaling in central surface density and radius, is the same for all. Moreover, this shape is inconsistent with the stars living in a NFW potential, likely because the corresponding three-dimensional stellar density profile has a central plateau or core (see Sect.~\ref{sec:intro}). Here we re-analyze the data to show how the equations derived in Sects.~\ref{sec:central_volume} and \ref{sec:shapes} may be used to determine the key parameter $\rho'(0)$.

\begin{figure*}[ht!] 
\centering
\includegraphics[width=0.66\linewidth]{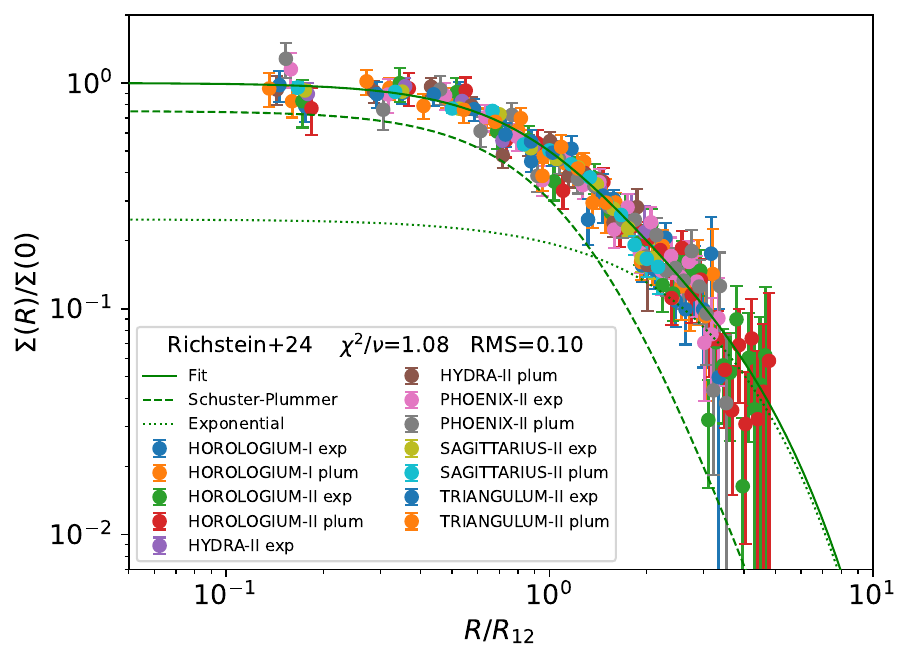}
\caption{
  The result of fitting Eq.~(\ref{eq:magic_fit}) to the UFDs  analyzed by \citet{2024ApJ...973L..15S}.  It is equivalent to Fig.~1 in \citet{2024ApJ...973L..15S} but with a different fitting function -- conforming to Eq.~(\ref{eq:sigma11}) in this case. The solid line shows the best fit whereas the other lines show the two terms contributing to the solution --   the surface density of a Schuster-Plummer function (the dashed line) and the surface density of an exponential profile (the dotted line). The values for $\Sigma(0)$ and $R_{12}$ used for scaling are taken from the best fit.
  \label{fig:read_richstein22}
}
\end{figure*}
Close to the center of the distribution, the surface density profiles from spherically symmetric volume densities have all the same dependence on radius, given by Eq.~(\ref{eq:sigma11}).  Thus,  $\rho'(0)$ may be inferred from Eq.~(\ref{eq:finally2}) by extrapolating to the center the derivatives of the observed $\Sigma(R)$. This exercise begins by fitting a single shape the surface density profiles of the six UFDs -- since each UFD surface density profile is inferred in two different ways \citep[see][]{2024ApJ...967...72R}, in the end we have twelve different surface density profiles to fit. The exercise was carried out by \citet{2024ApJ...973L..15S} but they do not use a functional form consistent with Eq.~(\ref{eq:sigma11}). Here we repeat it with a fitting function chosen to be  a linear combination of the surface densities corresponding to an exponential plus a Schuster-Plummer function (see Appendix~\ref{sec:appa}), namely,
\begin{equation}
  \begin{split}
  f(x) = 2C_0\,(x/C_1)\,K_1(x/C_1)+\\ C_2\frac{4/3}{\left[1+(x/C_3)^2\right]^2},
\end{split}
  \label{eq:magic_fit}
\end{equation}
so that the surface density of the $i$-th UFD  is
\begin{equation}
  \Sigma_i(R) = \Sigma_{i0}\,f(R/b_i),
  \label{eq:magic_fit2}
\end{equation}
with $\Sigma_{i0}$ and $b_i$ the coefficients that normalize each UFD to the same central density and radius.  Equation~(\ref{eq:magic_fit}) is consistent with Eq.~(\ref{eq:sigma11}) and retains the freedom to model $\rho'(0) \neq 0$, since the sum of the two surface density components satisfies Eq.~(\ref{eq:sigma11}), with one component permitting a non-zero $\rho'(0)$. The free parameter of the fits are $C_0,\dots C_3$ plus $b_i$ and $\Sigma_{i0}$, which makes a total of $4+12\times 2= 28$. With  207 observed points, this corresponds 179 degrees of freedom (hereafter $\nu$). The fit was carried out using the python routine {\tt least\_squares} from {\tt scipy.optimize} \citep{2020NatMe..17..261V}. The resulting fit is shown in Fig.~\ref{fig:read_richstein22} (the solid line). It is excellent, with a $\chi^2\simeq 1.08\,\nu$, meaning that the deviations from the fit are basically consistent with the error bars of the data, assigned as the Poisson error from star counting \citep[for details, see][]{2024ApJ...967...72R}. 

The fitted surface density profile shown in Fig.~\ref{fig:read_richstein22} is representative of the individual UFDs, rather than an artificial construct resulting from stacking multiple galaxies. This is demonstrated in Appendix~\ref{sec:appb}, where the global fit is overlaid on each surface density profile of the individual galaxy (see Fig.~\ref{fig:read_richstein22_}).

The red solid line in Fig.~\ref{fig:central_dip9_pub} shows $\rho'(r)$ corresponding to the profile in Fig.~\ref{fig:read_richstein22}, computed using the approximation in Eq.~(\ref{eq:finally2}). The formal uncertainties of this estimate are very small (the red shaded region), represented in Fig.~\ref{fig:central_dip9_pub}  as 2$\sigma$ error bars inferred by bootstrapping\footnote{The observed data were randomly resampled 200 times, and  the fit was repeated with each one of these mock profiles. The 2$\sigma$ error bars correspond to the band containing 95\,\% of all profiles around the best fit.}.  First, and more importantly, $\rho'(r)\to 0$ when $r\to 0$, which likely explains why the UFDs were found to be incompatible with cuspy NFW dark matter distribution in the analysis carried out by \citet{2024ApJ...973L..15S}. Note that the normalization used in Fig.~\ref{fig:central_dip9_pub} makes the estimate of $\rho'(r)$ independent of the central stellar density and global width of the stellar distribution, as we argue in Sec.~\ref{sec:central_volume}. Second, the details of the functional form used in Eq.~(\ref{eq:magic_fit}) do not seem to influence this result. We try other alternatives that also conform to Eq.~(\ref{eq:finally2}) and also result in $\rho'(0) \simeq  0$. Moreover,  the functional form used in Eq.~(\ref{eq:magic_fit}) is flexible enough to recover the correct $\rho'(r)$ from mock profiles that have both $\rho'(0)\not=0$ (Eq.~[\ref{eq:exp}])  and $\rho'(0)=0$ (Eq.~[\ref{eq:plummer}]). The latter follows from Monte Carlo simulations that mimic both the size and the error characteristics of the used UFD dataset. 

%
\section{Discussion and conclusions}\label{sec:discussion}

The inner slope of the three-dimensional stellar density distribution in dwarf galaxies ($\rho'[0]$) emerges as a key parameter for characterizing possible deviations of the true DM from the otherwise highly successful CDM paradigm. The reason can be traced to the incompatibility between a cored stellar distribution (Eq.~[\ref{eq:definition}]) and the gravitational potential produced by CDM model galaxies when stellar feedback is negligible (see Sect.~\ref{sec:intro}). Since the stellar distribution can be inferred from photometry alone, it is observationally inexpensive compared to the spectroscopic techniques commonly used to determine the structure of dark matter halos \citep[e.g.,][]{2008MNRAS.390...71C,2016CoPhC.200..336B,2019MNRAS.482.1525V}. While the standard spectroscopic approach is clearly preferred for bright galaxies when sufficient data can be obtained, it becomes a significant challenge for low-mass galaxies, which are precisely the systems needed to constrain the nature of DM. Thus, the alternative approach, based on the presence or absence of photometric cores, becomes particularly interesting, as it may allow the sample size to be increased enough to assess, with confidence, some of the claimed deviations from CDM (see Sect.~\ref{sec:intro}). 

The determination of whether the three-dimensional stellar distribution contains a core must rely on the observed, plane-of-the-sky projected surface density. Deprojecting real data, which are inevitably affected by noise, is highly ambiguous. Here, we work out a practical procedure to estimate the inner slope directly, thereby bypassing the need for an explicit deprojection. Under a set of reasonable assumptions (namely, spherical symmetry and a smooth finite stellar density profile), we find that the inner slope can be directly obtained from the radial derivatives of the observed surface density (Eq.~[\ref{eq:finally2}]). This formal expression is always valid, but numerically computing derivatives amplifies the noise in the data, calling its practical use into question. To address this issue, we have developed a workaround.  We demonstrate that the radial dependence of the stellar surface density corresponding to any volume density is always the same; see Eqs.~(\ref{eq:sigma11}) -- (\ref{eq:sigma11b}). By fitting the observed surface density to functions that include this functional form, $\Sigma(R)$ and its derivatives can be extrapolated to $R=0$, allowing $\rho'(0)$ to be estimated from Eq.~(\ref{eq:finally2}) while keeping noise effects minimal.

The technique has been tested on real data from UFD galaxies. These objects are particularly relevant in the present context, as they have been claimed to exhibit potential deviations from CDM \citep{2024ApJ...973L..15S}. In Sect.~\ref{sec:ufds}, we simultaneously fitted a set of six UFDs, each with two renderings, using a combination of a projected exponential and a projected Schuster–Plummer profile that satisfies Eq.~(\ref{eq:sigma11}), thereby retaining the flexibility to recover $\rho'(0) \neq 0$ (Eq.~[\ref{eq:magic_fit}]). All galaxies are assumed to have a density profile with the same shape but with a different scaling in central density and radius. The resulting fit is shown in Fig.~\ref{fig:read_richstein22}, which is very good since $\chi^2/\nu\sim 1.08$. Importantly, the fit works equally well for each individual UFD, confirming that all UFDs share the same shape (see Fig.~\ref{fig:read_richstein22_} in Appendix~\ref{sec:appb}). The extrapolation to the center of the fitted profile yields $\rho'(0)\simeq 0$, as shown by the red line and region in Fig.~\ref{fig:central_dip9_pub}. The existence of such stellar core explains the incompatibility of these profiles with cuspy NFW potentials pointed out by \citet{2024ApJ...973L..15S}.

Some of the individual UFD surface densities may hint at the presence of a central dip (see, e.g., Horologium I and II in Fig.~\ref{fig:read_richstein22_}, Appendix~\ref{sec:appb}). If dips like these were confirmed to be real rather than a statistical fluctuation, they may be extremely telling from a physical stand point. In our parlance, the dip implies $d\Sigma(R)/dR > 0$ at some small but non-zero $R$.  Because of Eq.~(\ref{eq:deriv1}), the dip requires $\rho'(r)$ to be positive over a range of radii. Using Eq.~(\ref{eq:first_derivative}), and keeping in mind that $I'(R_0)$ is negative (see Eq.~[\ref{eq:the_tail}]),   $d\Sigma(R)/dR > 0$ implies $\rho'(r) > 0$ at the center. Thus, for the dip to exist, the volume density of stars must have a central depression, dubbed here as {\em central hole}. This central hole could be caused by a number of very interesting physical mechanisms, including core scouring by several massive black holes \citep[e.g.,][]{2010MNRAS.407..447H,2021MNRAS.502.4794N}, the expansion of the dark matter halo driven by self-interactions or other physics beyond the standard CDM framework \citep[e.g.,][]{2025A&A...702A.113S}, or it may indicate that gravity is governed by MOND (MOdified Newtonian Dynamics; \citealt{1983ApJ...270..365M}) rather than by Newtonian gravity \citep{2022ApJ...940...46S}. We bring this up here because the above formalism allows the retrieval of a volume density gradient of any sign. In particular, when equipped with the appropriate approximations (Eq.~[\ref{eq:sigma11}]), it enables the measurement of $\rho'(r) > 0$. Thus, the tools developed here may offer a way to explore the exotic physical mechanisms mentioned above.
 
Two further results and conclusions deserve explicit mention:
(a) The scaling used to display the estimate of $\rho'(0)$ in Fig.~\ref{fig:central_dip9_pub} is based entirely on observational data and is independent of any shifts in the amplitude or radial extent of the observed surface density.
(b) When $R \to 0$, the surface density $\Sigma(R)$ scales as $R^2 \ln R$ if $\rho'(0) \neq 0$, or as $R^2$ if $\rho'(0) = 0$ (see Eqs.~[\ref{eq:sigma11}]–[\ref{eq:sigma11b}]). This implies that $d\Sigma/dR \to 0$ as $R \to 0$, so that $\Sigma(R)$ always exhibits a central plateau for any finite $\rho(r)$. Consequently, information about the value of $\rho'(0)$ is encoded in the way $\Sigma(R)$ declines from this plateau. The mentioned approximation is accurate to within one percent for $R \lesssim R_{12}$ (Fig.~\ref{fig:central_dip8}).

\begin{acknowledgments}
This paper is dedicated to the memory of the late Prof. Francisco S\'anchez, a pioneer of modern Spanish astrophysics and my father.
The work is based on inspiring conversations with Ignacio Trujillo on how to interpret surface density profiles with a slight inner upturn, that may have been observed in some profiles \citep[e.g., {\em Nube};][]{2024A&A...681A..15M}.
Special thanks are due to an anonymous referee who provided more rigorous derivations for  Eqs.~(\ref{eq:finally2_old}) and (\ref{eq:sigma11}). In particular, the derivation of  Eq.~(\ref{eq:sigma11}) in Appendix~\ref{sec:appc} follows an idea proposed by the referee.
We acknowledge financial support from the Spanish Ministry of Science and Innovation, project PID2022-136598NB-C31 (ESTALLIDOS8) and from the EU UNDARK project (project number 101159929). 
\end{acknowledgments}

%

\vspace{5mm}


\software{
          SciPy \citep{2020NatMe..17..261V}
          }



\appendix

\section{Surface density and its derivatives for two analytical volume densities}\label{sec:appa}


For an exponential volume density profile (Eq.~[\ref{eq:exp}] with $\rho_s=r_s=1$), the corresponding surface density is
\begin{equation}
\Sigma(R) =2\,R\,K_1(R),
\end{equation}
with $K_1$ the modified Bessel function of the second kind of order one \citep{1965tisp.book.....G}.  The first derivative of $K_1$ is 
\begin{equation}
K'_1(R) =-\frac{1}{2}\left[K_0(R)+K_2(R)\right],
\end{equation}
since there are general recurrent equations for the derivatives of these functions \citep{1972hmfw.book.....A}. Thus, in this case,
\begin{equation}
  \Sigma'(R)= - R K_0(R)+ 2K_1(R)- R K_2(R).
  \label{eq:deriv_exp}
\end{equation}
and
\begin{equation}
\Sigma''(R)= 
\frac{R}{2}  K_{-1}(R)-2K_0(R)+R K_1(R)-2K_2(R)+\frac{R}{2}K_3(R).
\end{equation}

For a Schuster-Plummer profile (Eq.~[\ref{eq:plummer}] with $\rho_s=r_s=1$), the corresponding surface density is \citep[e.g.,][]{2022Univ....8..214S}
\begin{equation}
\Sigma(R) =\frac{4/3}{(1+R^2)^2},
\end{equation}
with
\begin{equation}
  \Sigma'(R) = -\frac{16\,R}{3\,(1+R^2)^3},
\end{equation}
and
\begin{equation}
  \Sigma''(R) =
  \frac{32\,R^2}{(1+R^2)^4}-\frac{16}{3\,(1+R^2)^3}.
\end{equation}

%
\newcommand{\eqsigma}{\ref{eq:sigma11}}
\section{Alternative derivation of Eq.~(\eqsigma )}\label{sec:appc}
Equation~(\ref{eq:finally2_old}) takes also the form of
\begin{equation}
  \frac{d}{dR}\left[\frac{\Sigma'(R)}{R}\right] \simeq -2 \frac{\rho'(R)}{R},
  \label{eq:finally2_super}
\end{equation}
which can be integrated expanding de volume density around zero, $\rho'(R)\simeq \rho'(0)+\rho''(0) R+\dots$, so that
\begin{equation}
\frac{\Sigma'(R)}{R}\simeq P_1-2 \rho'(0) \ln R - 2 \rho''(0) R+\dots ,
\end{equation}
with $P_1$ a constant of integration. A second integration yields,
\begin{equation}
\Sigma(R) \simeq \Sigma(0)-2\rho'(0) \frac{R^2(2\ln R-1)}{4}+\frac{P_1}{2} R^2-\frac{2}{3} \rho''(0) R^3+\dots,
\end{equation}
or, equivalently,
\begin{equation}
  \Sigma(R)\simeq \Sigma(0)-\rho'(0)\, R^2\,\ln R+ \frac{1}{2}\left[P_1+\rho'(0)\right]\, R^2-\frac{2}{3} \rho''(0) R^3+\dots,
  \label{eq:newsigma11}
\end{equation}
which is Eq.~(\ref{eq:sigma11}) showing an additional $R^3$ term in the expansion. A term-by-term comparison between Eqs.~(\ref{eq:sigma11b}) and (\ref{eq:newsigma11}) gives the equivalence 
\begin{equation}
   \frac{P_1}{2} = \rho'(0)\,\ln(2R_0)+I'(R_0),
\end{equation}
showing that the integration constant $P_1$ encodes the information associated with the splitting radius $R_0$ and the integral $I'(R_0)$ used in the main text. Because of Eq.~(\ref{eq:finally2_super}),
\begin{equation}
\frac{1}{2}\left[\Sigma'(R)/R-\Sigma''(R)\right]\simeq \rho'(0)+\rho''(0)\, R+\dots .
\end{equation}

%
\section{Surface density profile fit on the individual UFD data}\label{sec:appb}
\begin{figure*}[ht!] 
  \centering
  \begin{minipage}{0.75\linewidth} 
    \centering
    \includegraphics[width=0.48\linewidth]{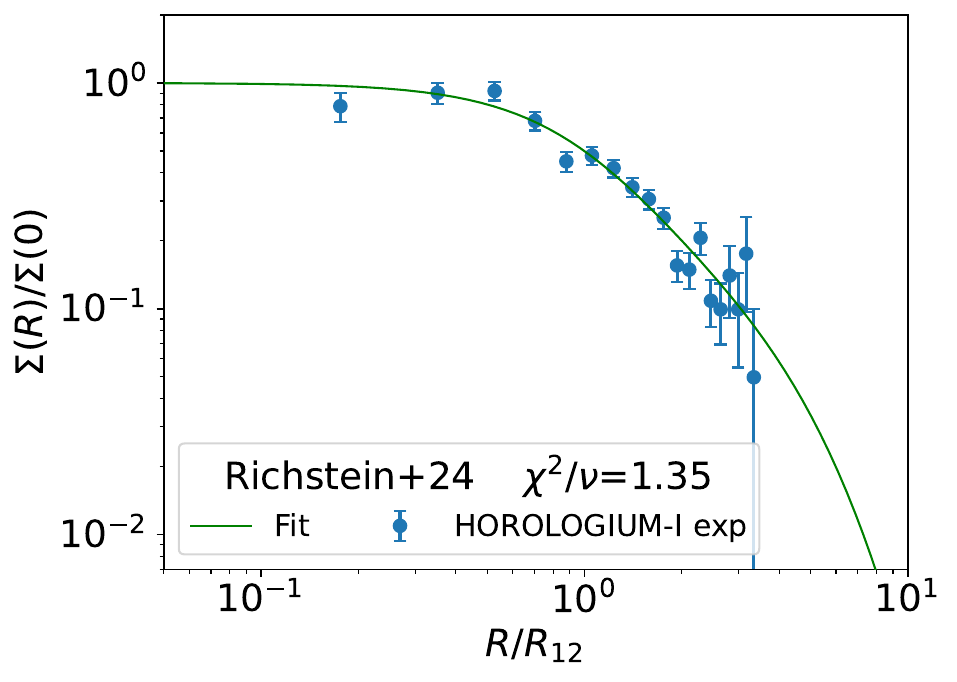}
    \includegraphics[width=0.48\linewidth]{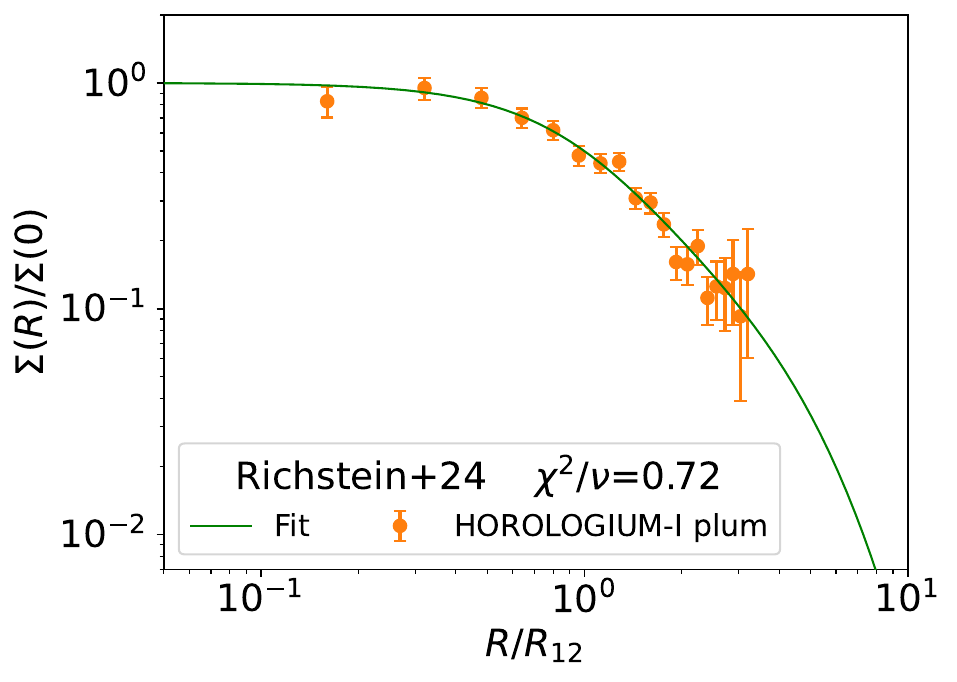}\\
    \includegraphics[width=0.48\linewidth]{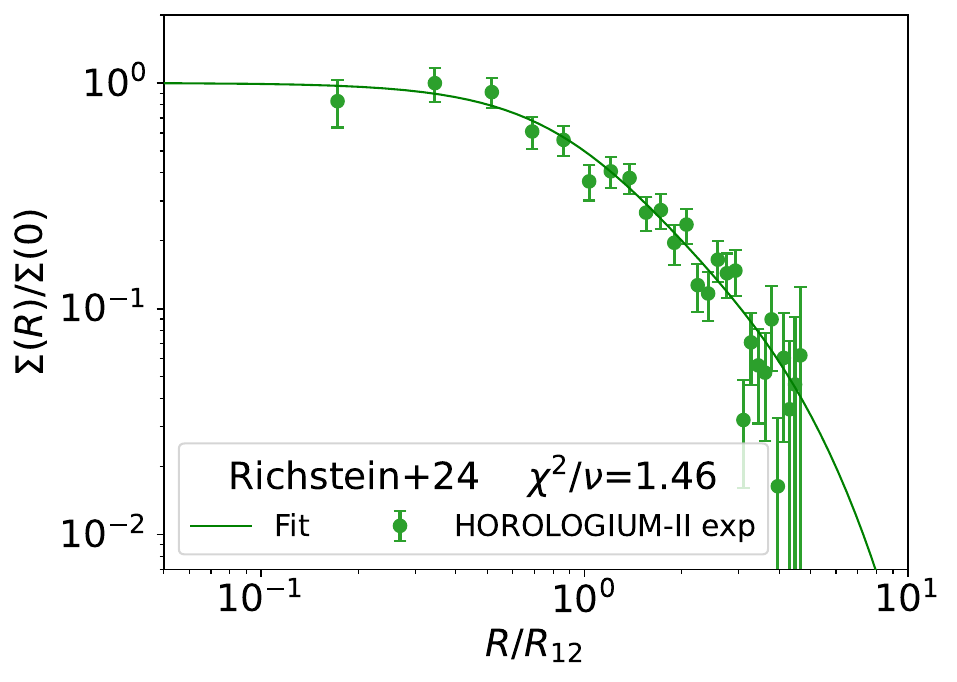}
    \includegraphics[width=0.48\linewidth]{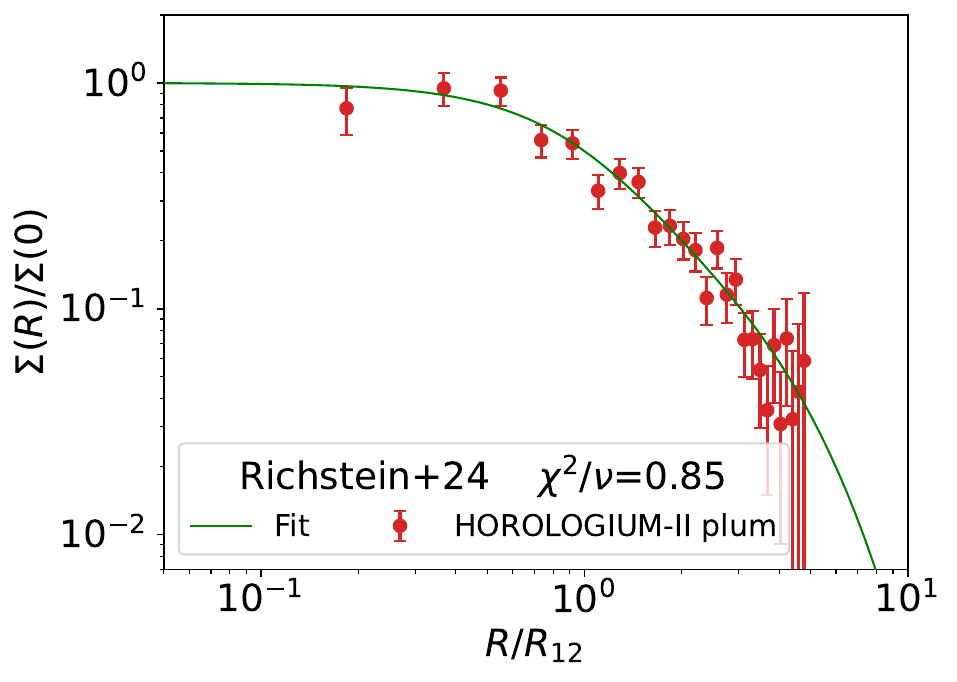}\\
    \includegraphics[width=0.48\linewidth]{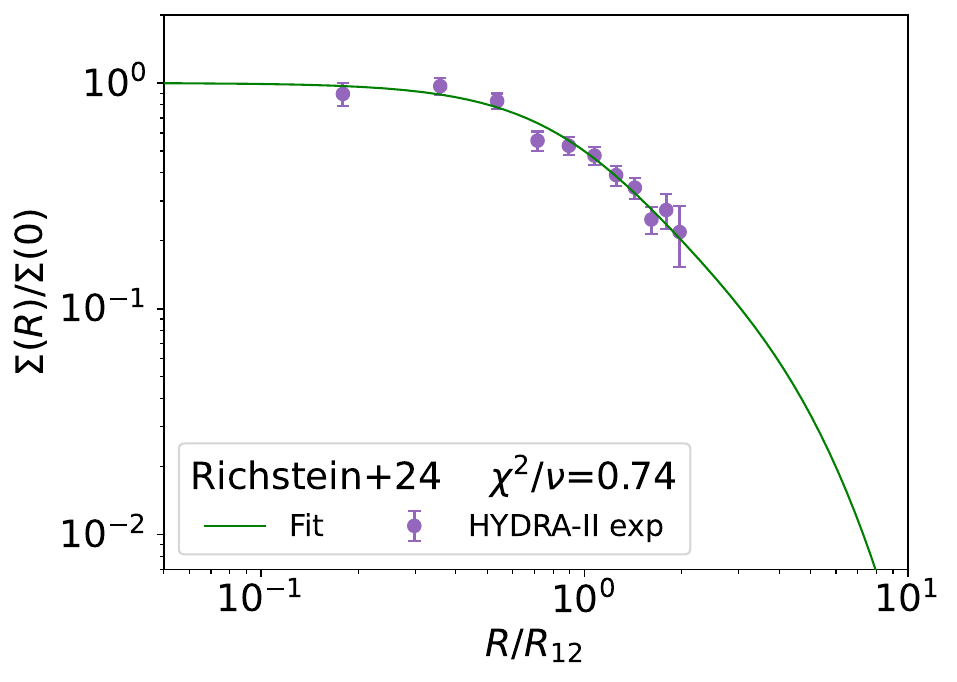}
    \includegraphics[width=0.48\linewidth]{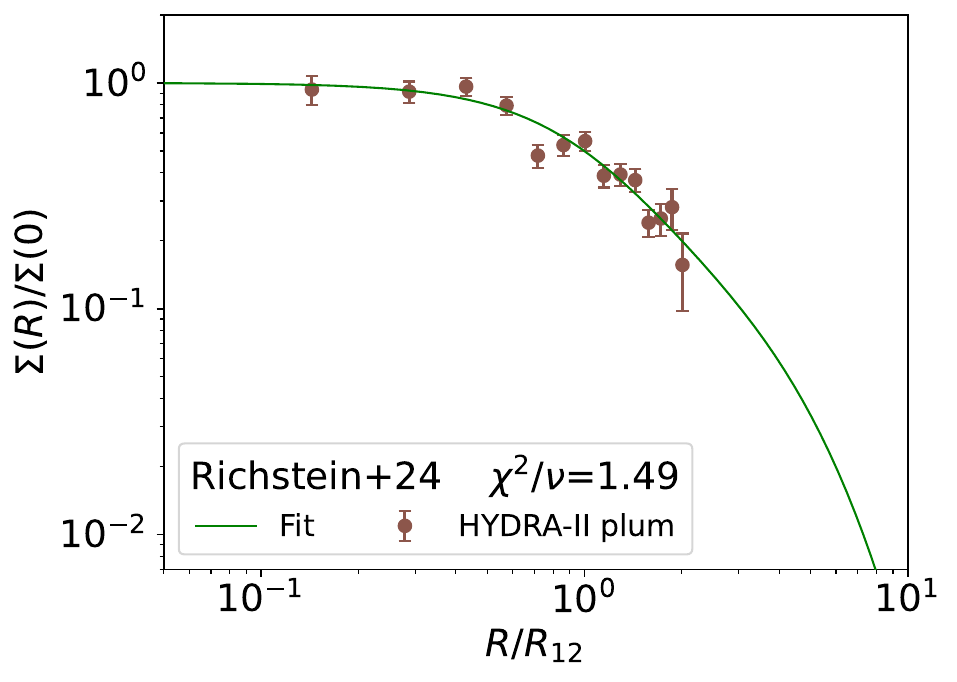}\\
    \includegraphics[width=0.48\linewidth]{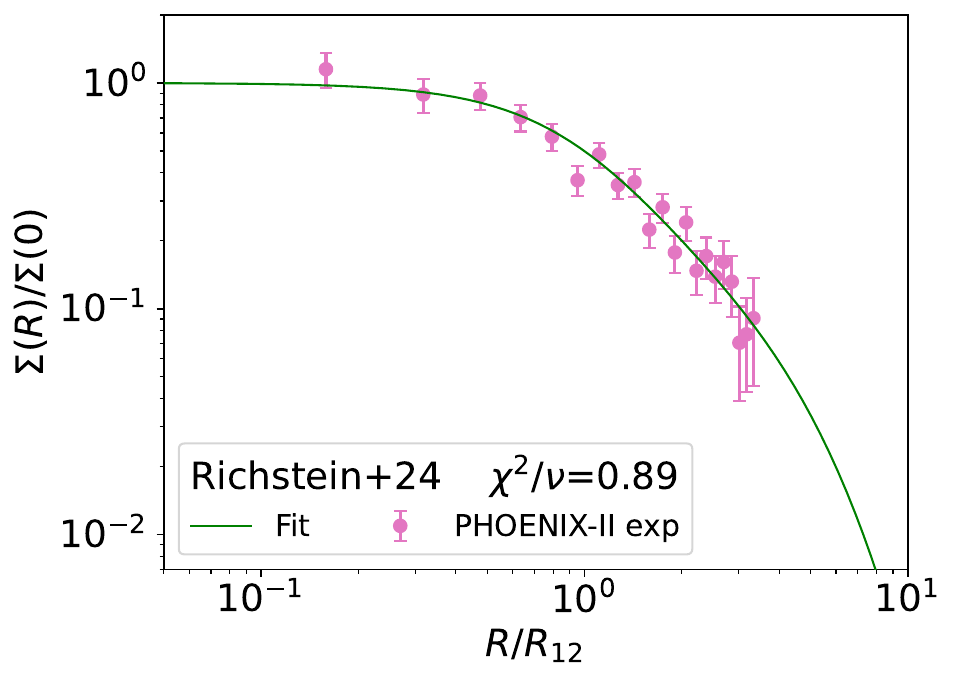}
    \includegraphics[width=0.48\linewidth]{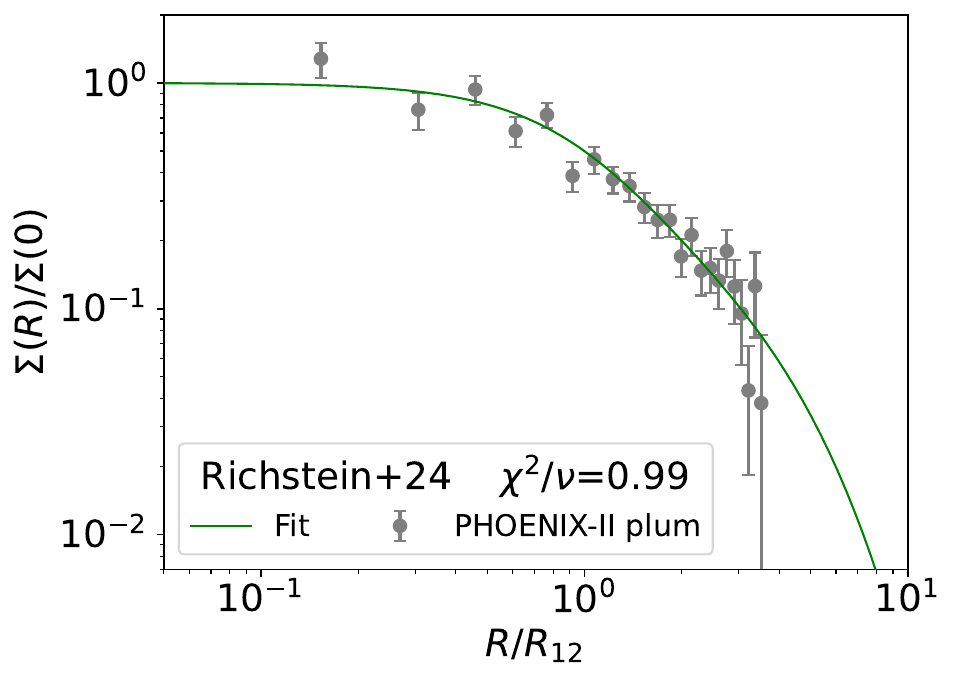}\\
    \includegraphics[width=0.48\linewidth]{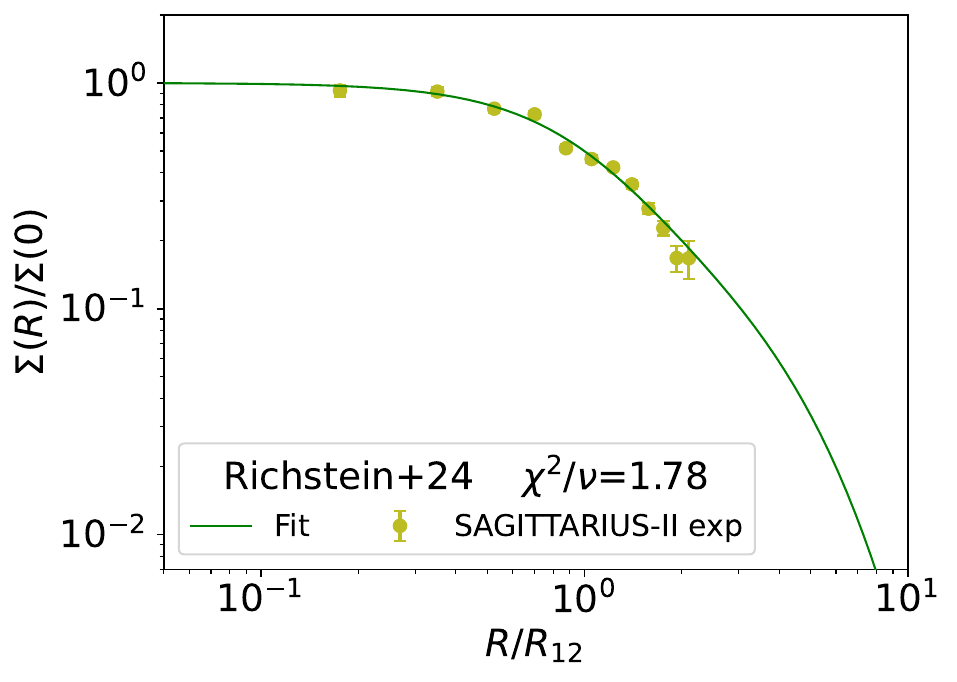}
    \includegraphics[width=0.48\linewidth]{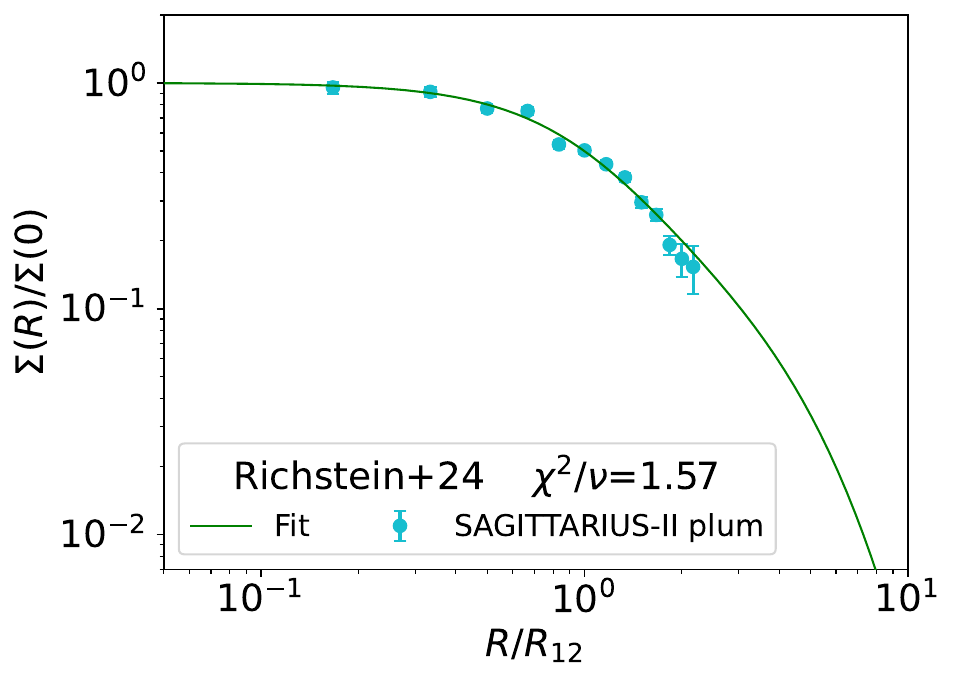}\\
    \includegraphics[width=0.48\linewidth]{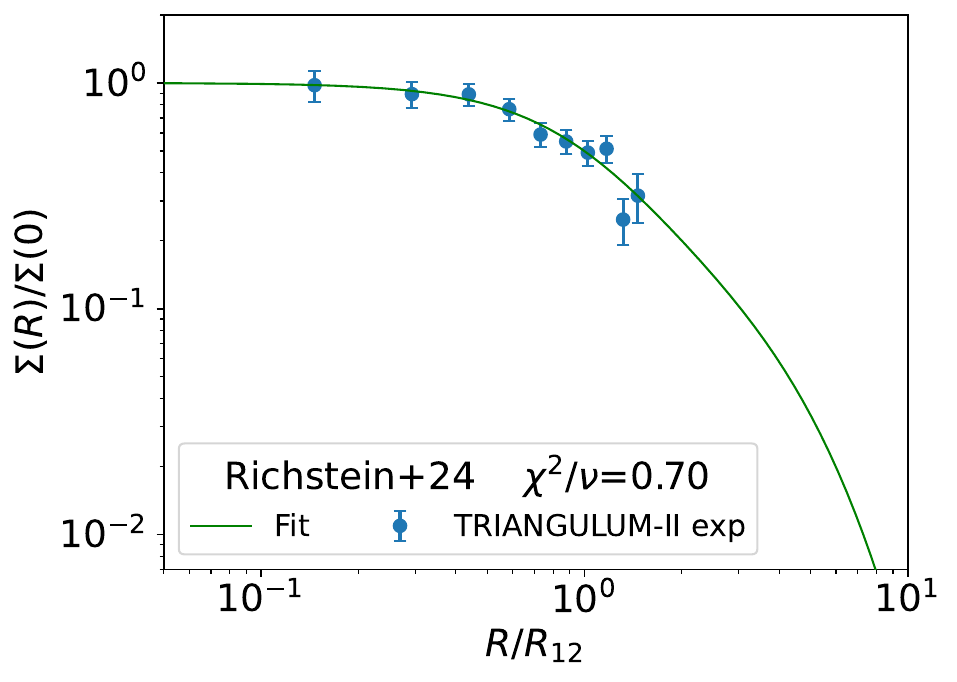}
    \includegraphics[width=0.48\linewidth]{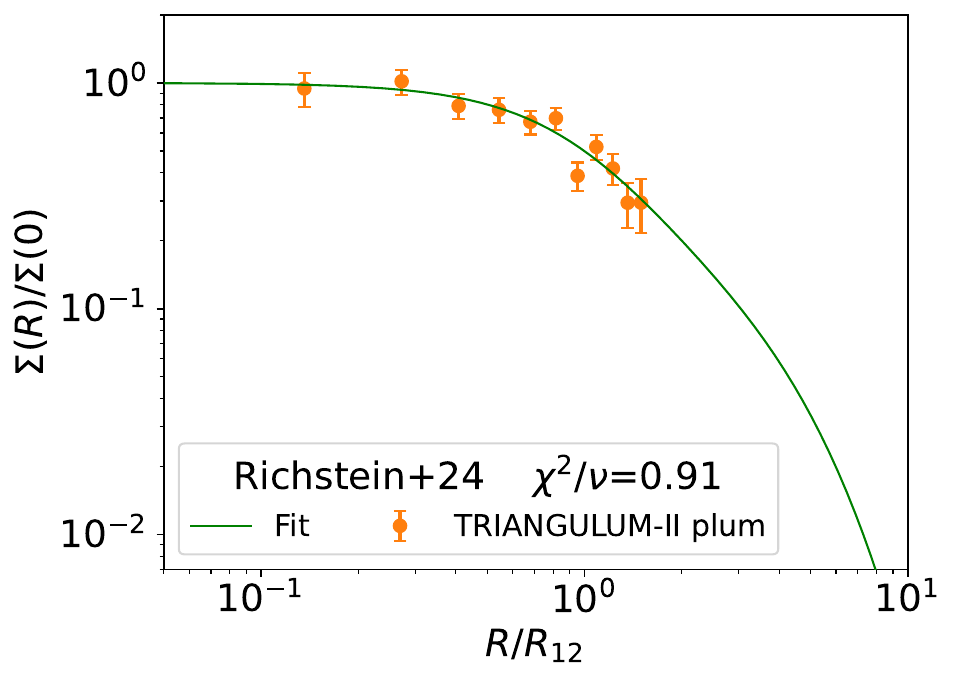}
  \end{minipage}
  \hfill
  \begin{minipage}{0.20\linewidth}
    \vspace{12cm}
    \caption{
      Similar to Fig.~\ref{fig:read_richstein22} but showing each UFD profile individually. The green solid line in all panels is the same, and the color code is also copied  from Fig.~\ref{fig:read_richstein22}. Note how the fit provided by Eq.~(\ref{eq:magic_fit}) is not only good for the full set but is good for each individual galaxy. Note also that some galaxies may show a central drop in surface density (e.g., Horologium I and II), although for the moment is just a fluctuation within error bars. The name of each object and its $\chi^2/\nu$ is shown in the individual insets.
      \label{fig:read_richstein22_}
    }
  \end{minipage}
\end{figure*}

Here we overlie on the observed profile of the individual UFDs the fit to all of them together worked out in Sect.~\ref{sec:ufds} and shown in Fig.~\ref{fig:read_richstein22}. Figure~\ref{fig:read_richstein22_} represents the common fit (the green solid line) on the data of the individual UFD (note that each UFD is shown twice since they have two profiles constructed from the raw data with a slightly different approach -- see \citeauthor{2024ApJ...967...72R}~\citeyear{2024ApJ...967...72R} for details). The figure shows how the fit provided by Eq.~(\ref{eq:magic_fit}) is not only good for the full set but is good for each individual galaxy. In other words, the fit in Fig.~\ref{fig:read_richstein22} is representative of the individual UFDs, rather than being an artificial construct produced by the stacking of different profiles.
%
%



\begin{thebibliography}{}
\expandafter\ifx\csname natexlab\endcsname\relax\def\natexlab#1{#1}\fi
\providecommand{\url}[1]{\href{#1}{#1}}
\providecommand{\dodoi}[1]{doi:~\href{http://doi.org/#1}{\nolinkurl{#1}}}
\providecommand{\doeprint}[1]{\href{http://ascl.net/#1}{\nolinkurl{http://ascl.net/#1}}}
\providecommand{\doarXiv}[1]{\href{https://arxiv.org/abs/#1}{\nolinkurl{https://arxiv.org/abs/#1}}}

\bibitem[{{Abramowitz} \& {Stegun}(1972)}]{1972hmfw.book.....A}
{Abramowitz}, M., \& {Stegun}, I.~A. 1972, {Handbook of Mathematical Functions}
  (Washington: US Department od Commerce, NBS)

\bibitem[{{An} \& {Evans}(2006)}]{2006ApJ...642..752A}
{An}, J.~H., \& {Evans}, N.~W. 2006, \apj, 642, 752, \dodoi{10.1086/501040}

\bibitem[{{Bonnivard} {et~al.}(2016){Bonnivard}, {H{\"u}tten}, {Nezri},
  {Charbonnier}, {Combet}, \& {Maurin}}]{2016CoPhC.200..336B}
{Bonnivard}, V., {H{\"u}tten}, M., {Nezri}, E., {et~al.} 2016, Computer Physics
  Communications, 200, 336, \dodoi{10.1016/j.cpc.2015.11.012}

\bibitem[{{Bullock} \& {Boylan-Kolchin}(2017)}]{2017ARA&A..55..343B}
{Bullock}, J.~S., \& {Boylan-Kolchin}, M. 2017, \araa, 55, 343,
  \dodoi{10.1146/annurev-astro-091916-055313}

\bibitem[{{Cappellari}(2008)}]{2008MNRAS.390...71C}
{Cappellari}, M. 2008, \mnras, 390, 71,
  \dodoi{10.1111/j.1365-2966.2008.13754.x}

\bibitem[{{Ciotti}(1999)}]{1999ApJ...520..574C}
{Ciotti}, L. 1999, \apj, 520, 574, \dodoi{10.1086/307478}

\bibitem[{{Ciotti} \& {Morganti}(2010)}]{2010MNRAS.408.1070C}
{Ciotti}, L., \& {Morganti}, L. 2010, \mnras, 408, 1070,
  \dodoi{10.1111/j.1365-2966.2010.17184.x}

\bibitem[{{Ciotti} \& {Pellegrini}(1992)}]{1992MNRAS.255..561C}
{Ciotti}, L., \& {Pellegrini}, S. 1992, \mnras, 255, 561,
  \dodoi{10.1093/mnras/255.4.561}

\bibitem[{{de Vaucouleurs}(1948)}]{1948AnAp...11..247D}
{de Vaucouleurs}, G. 1948, Annales d'Astrophysique, 11, 247

\bibitem[{{Del Popolo} \& {Le Delliou}(2021)}]{2021Galax...9..123D}
{Del Popolo}, A., \& {Le Delliou}, M. 2021, Galaxies, 9,
  \dodoi{10.3390/galaxies9040123}

\bibitem[{{Gradshteyn} \& {Ryzhik}(1965)}]{1965tisp.book.....G}
{Gradshteyn}, I.~S., \& {Ryzhik}, I.~M. 1965, {Table of integrals, series and
  products} (San Diego, CA: Academic Press)

\bibitem[{{Hernquist}(1990)}]{1990ApJ...356..359H}
{Hernquist}, L. 1990, \apj, 356, 359, \dodoi{10.1086/168845}

\bibitem[{{Hopkins} \& {Hernquist}(2010)}]{2010MNRAS.407..447H}
{Hopkins}, P.~F., \& {Hernquist}, L. 2010, \mnras, 407, 447,
  \dodoi{10.1111/j.1365-2966.2010.16915.x}

\bibitem[{{Ivezi{\'c}} {et~al.}(2019){Ivezi{\'c}}, {Kahn}, {Tyson}, {Abel},
  {Acosta}, {Allsman}, {Alonso}, {AlSayyad}, {Anderson}, {Andrew}, {Angel},
  {Angeli}, {Ansari}, {Antilogus}, {Araujo}, {Armstrong}, {Arndt}, {Astier},
  {Aubourg}, {Auza}, {Axelrod}, {Bard}, {Barr}, {Barrau}, {Bartlett}, {Bauer},
  {Bauman}, {Baumont}, {Bechtol}, {Bechtol}, {Becker}, {Becla}, {Beldica},
  {Bellavia}, {Bianco}, {Biswas}, {Blanc}, {Blazek}, {Blandford}, {Bloom},
  {Bogart}, {Bond}, {Booth}, {Borgland}, {Borne}, {Bosch}, {Boutigny},
  {Brackett}, {Bradshaw}, {Brandt}, {Brown}, {Bullock}, {Burchat}, {Burke},
  {Cagnoli}, {Calabrese}, {Callahan}, {Callen}, {Carlin}, {Carlson},
  {Chandrasekharan}, {Charles-Emerson}, {Chesley}, {Cheu}, {Chiang}, {Chiang},
  {Chirino}, {Chow}, {Ciardi}, {Claver}, {Cohen-Tanugi}, {Cockrum}, {Coles},
  {Connolly}, {Cook}, {Cooray}, {Covey}, {Cribbs}, {Cui}, {Cutri}, {Daly},
  {Daniel}, {Daruich}, {Daubard}, {Daues}, {Dawson}, {Delgado}, {Dellapenna},
  {de Peyster}, {de Val-Borro}, {Digel}, {Doherty}, {Dubois},
  {Dubois-Felsmann}, {Durech}, {Economou}, {Eifler}, {Eracleous}, {Emmons},
  {Fausti Neto}, {Ferguson}, {Figueroa}, {Fisher-Levine}, {Focke}, {Foss},
  {Frank}, {Freemon}, {Gangler}, {Gawiser}, {Geary}, {Gee}, {Geha}, {Gessner},
  {Gibson}, {Gilmore}, {Glanzman}, {Glick}, {Goldina}, {Goldstein}, {Goodenow},
  {Graham}, {Gressler}, {Gris}, {Guy}, {Guyonnet}, {Haller}, {Harris},
  {Hascall}, {Haupt}, {Hernandez}, {Herrmann}, {Hileman}, {Hoblitt}, {Hodgson},
  {Hogan}, {Howard}, {Huang}, {Huffer}, {Ingraham}, {Innes}, {Jacoby}, {Jain},
  {Jammes}, {Jee}, {Jenness}, {Jernigan}, {Jevremovi{\'c}}, {Johns}, {Johnson},
  {Johnson}, {Jones}, {Juramy-Gilles}, {Juri{\'c}}, {Kalirai}, {Kallivayalil},
  {Kalmbach}, {Kantor}, {Karst}, {Kasliwal}, {Kelly}, {Kessler}, {Kinnison},
  {Kirkby}, {Knox}, {Kotov}, {Krabbendam}, {Krughoff}, {Kub{\'a}nek},
  {Kuczewski}, {Kulkarni}, {Ku}, {Kurita}, {Lage}, {Lambert}, {Lange},
  {Langton}, {Le Guillou}, {Levine}, {Liang}, {Lim}, {Lintott}, {Long},
  {Lopez}, {Lotz}, {Lupton}, {Lust}, {MacArthur}, {Mahabal}, {Mandelbaum},
  {Markiewicz}, {Marsh}, {Marshall}, {Marshall}, {May}, {McKercher}, {McQueen},
  {Meyers}, {Migliore}, {Miller}, \& {Mills}}]{2019ApJ...873..111I}
{Ivezi{\'c}}, {\v{Z}}., {Kahn}, S.~M., {Tyson}, J.~A., {et~al.} 2019, \apj,
  873, 111, \dodoi{10.3847/1538-4357/ab042c}

\bibitem[{{Laureijs} {et~al.}(2011){Laureijs}, {Amiaux}, {Arduini},
  {Augu{\`e}res}, {Brinchmann}, {Cole}, {Cropper}, {Dabin}, {Duvet}, {Ealet},
  {Garilli}, {Gondoin}, {Guzzo}, {Hoar}, {Hoekstra}, {Holmes}, {Kitching},
  {Maciaszek}, {Mellier}, {Pasian}, {Percival}, {Rhodes}, {Saavedra Criado},
  {Sauvage}, {Scaramella}, {Valenziano}, {Warren}, {Bender}, {Castander},
  {Cimatti}, {Le F{\`e}vre}, {Kurki-Suonio}, {Levi}, {Lilje}, {Meylan},
  {Nichol}, {Pedersen}, {Popa}, {Rebolo Lopez}, {Rix}, {Rottgering},
  {Zeilinger}, {Grupp}, {Hudelot}, {Massey}, {Meneghetti}, {Miller}, {Paltani},
  {Paulin-Henriksson}, {Pires}, {Saxton}, {Schrabback}, {Seidel}, {Walsh},
  {Aghanim}, {Amendola}, {Bartlett}, {Baccigalupi}, {Beaulieu}, {Benabed},
  {Cuby}, {Elbaz}, {Fosalba}, {Gavazzi}, {Helmi}, {Hook}, {Irwin}, {Kneib},
  {Kunz}, {Mannucci}, {Moscardini}, {Tao}, {Teyssier}, {Weller}, {Zamorani},
  {Zapatero Osorio}, {Boulade}, {Foumond}, {Di Giorgio}, {Guttridge}, {James},
  {Kemp}, {Martignac}, {Spencer}, {Walton}, {Bl{\"u}mchen}, {Bonoli},
  {Bortoletto}, {Cerna}, {Corcione}, {Fabron}, {Jahnke}, {Ligori}, {Madrid},
  {Martin}, {Morgante}, {Pamplona}, {Prieto}, {Riva}, {Toledo}, {Trifoglio},
  {Zerbi}, {Abdalla}, {Douspis}, {Grenet}, {Borgani}, {Bouwens}, {Courbin},
  {Delouis}, {Dubath}, {Fontana}, {Frailis}, {Grazian}, {Koppenh{\"o}fer},
  {Mansutti}, {Melchior}, {Mignoli}, {Mohr}, {Neissner}, {Noddle}, {Poncet},
  {Scodeggio}, {Serrano}, {Shane}, {Starck}, {Surace}, {Taylor},
  {Verdoes-Kleijn}, {Vuerli}, {Williams}, {Zacchei}, {Altieri}, {Escudero
  Sanz}, {Kohley}, {Oosterbroek}, {Astier}, {Bacon}, {Bardelli}, {Baugh},
  {Bellagamba}, {Benoist}, {Bianchi}, {Biviano}, {Branchini}, {Carbone},
  {Cardone}, {Clements}, {Colombi}, {Conselice}, {Cresci}, {Deacon}, {Dunlop},
  {Fedeli}, {Fontanot}, {Franzetti}, {Giocoli}, {Garcia-Bellido}, {Gow},
  {Heavens}, {Hewett}, {Heymans}, {Holland}, {Huang}, {Ilbert}, {Joachimi},
  {Jennins}, {Kerins}, {Kiessling}, {Kirk}, {Kotak}, {Krause}, {Lahav}, {van
  Leeuwen}, {Lesgourgues}, {Lombardi}, {Magliocchetti}, {Maguire}, {Majerotto},
  {Maoli}, {Marulli}, {Maurogordato}, {McCracken}, {McLure}, {Melchiorri},
  {Merson}, {Moresco}, {Nonino}, {Norberg}, {Peacock}, {Pello}, {Penny},
  {Pettorino}, {Di Porto}, {Pozzetti}, {Quercellini}, {Radovich}, {Rassat},
  {Roche}, {Ronayette}, \& {Rossetti}}]{2011arXiv1110.3193L}
{Laureijs}, R., {Amiaux}, J., {Arduini}, S., {et~al.} 2011, arXiv e-prints,
  arXiv:1110.3193, \dodoi{10.48550/arXiv.1110.3193}

\bibitem[{{Lynden-Bell}(1962)}]{1962MNRAS.123..447L}
{Lynden-Bell}, D. 1962, \mnras, 123, 447, \dodoi{10.1093/mnras/123.5.447}

\bibitem[{{Milgrom}(1983)}]{1983ApJ...270..365M}
{Milgrom}, M. 1983, \apj, 270, 365, \dodoi{10.1086/161130}

\bibitem[{{Montes} {et~al.}(2024){Montes}, {Trujillo}, {Karunakaran},
  {Infante-Sainz}, {Spekkens}, {Golini}, {Beasley}, {Cebri{\'a}n}, {Chamba},
  {D'Onofrio}, {Kelvin}, \& {Rom{\'a}n}}]{2024A&A...681A..15M}
{Montes}, M., {Trujillo}, I., {Karunakaran}, A., {et~al.} 2024, \aap, 681, A15,
  \dodoi{10.1051/0004-6361/202347667}

\bibitem[{{Nasim} {et~al.}(2021){Nasim}, {Gualandris}, {Read}, {Antonini},
  {Dehnen}, \& {Delorme}}]{2021MNRAS.502.4794N}
{Nasim}, I.~T., {Gualandris}, A., {Read}, J.~I., {et~al.} 2021, \mnras, 502,
  4794, \dodoi{10.1093/mnras/stab435}

\bibitem[{{Navarro} {et~al.}(1997){Navarro}, {Frenk}, \&
  {White}}]{1997ApJ...490..493N}
{Navarro}, J.~F., {Frenk}, C.~S., \& {White}, S. D.~M. 1997, \apj, 490, 493,
  \dodoi{10.1086/304888}

\bibitem[{{Pe{\~n}arrubia} {et~al.}(2012){Pe{\~n}arrubia}, {Pontzen}, {Walker},
  \& {Koposov}}]{2012ApJ...759L..42P}
{Pe{\~n}arrubia}, J., {Pontzen}, A., {Walker}, M.~G., \& {Koposov}, S.~E. 2012,
  \apjl, 759, L42, \dodoi{10.1088/2041-8205/759/2/L42}

\bibitem[{{Richstein} {et~al.}(2024){Richstein}, {Kallivayalil}, {Simon},
  {Garling}, {Wetzel}, {Warfield}, {van der Marel}, {Jeon}, {Rose}, {Torrey},
  {Engelhardt}, {Besla}, {Choi}, {Geha}, {Guhathakurta}, {Kirby}, {Patel},
  {Sacchi}, \& {Sohn}}]{2024ApJ...967...72R}
{Richstein}, H., {Kallivayalil}, N., {Simon}, J.~D., {et~al.} 2024, \apj, 967,
  72, \dodoi{10.3847/1538-4357/ad393c}

\bibitem[{{S{\'a}nchez Almeida}(2022{\natexlab{a}})}]{2022ApJ...940...46S}
{S{\'a}nchez Almeida}, J. 2022{\natexlab{a}}, \apj, 940, 46,
  \dodoi{10.3847/1538-4357/ac9520}

\bibitem[{{S{\'a}nchez Almeida}(2022{\natexlab{b}})}]{2022Univ....8..214S}
---. 2022{\natexlab{b}}, Universe, 8, 214, \dodoi{10.3390/universe8040214}

\bibitem[{{S{\'a}nchez Almeida}(2024)}]{2024RNAAS...8..167S}
---. 2024, Research Notes of the American Astronomical Society, 8

\bibitem[{{S{\'a}nchez Almeida} {et~al.}(2023){S{\'a}nchez Almeida},
  {Plastino}, \& {Trujillo}}]{2023ApJ...954..153S}
{S{\'a}nchez Almeida}, J., {Plastino}, A.~R., \& {Trujillo}, I. 2023, \apj,
  954, 153, \dodoi{10.3847/1538-4357/ace534}

\bibitem[{{S{\'a}nchez Almeida} {et~al.}(2024{\natexlab{a}}){S{\'a}nchez
  Almeida}, {Plastino}, \& {Trujillo}}]{2024A&A...690A.151S}
---. 2024{\natexlab{a}}, \aap, 690, A151, \dodoi{10.1051/0004-6361/202449187}

\bibitem[{{S{\'a}nchez Almeida} {et~al.}(2025){S{\'a}nchez Almeida},
  {Plastino}, \& {Trujillo}}]{2025A&A...702A.113S}
---. 2025, \aap, 702, A113, \dodoi{10.1051/0004-6361/202555679}

\bibitem[{{S{\'a}nchez Almeida} {et~al.}(2024{\natexlab{b}}){S{\'a}nchez
  Almeida}, {Trujillo}, \& {Plastino}}]{2024ApJ...973L..15S}
{S{\'a}nchez Almeida}, J., {Trujillo}, I., \& {Plastino}, A.~R.
  2024{\natexlab{b}}, \apjl, 973, L15, \dodoi{10.3847/2041-8213/ad66bc}

\bibitem[{{Tollet} {et~al.}(2016){Tollet}, {Macci{\`o}}, {Dutton}, {Stinson},
  {Wang}, {Penzo}, {Gutcke}, {Buck}, {Kang}, {Brook}, {Di Cintio}, {Keller}, \&
  {Wadsley}}]{2016MNRAS.456.3542T}
{Tollet}, E., {Macci{\`o}}, A.~V., {Dutton}, A.~A., {et~al.} 2016, \mnras, 456,
  3542, \dodoi{10.1093/mnras/stv2856}

\bibitem[{{Valenciano} {et~al.}(2025){Valenciano}, {Camalich}, {Di Cintio},
  {Navarro}, {Battaglia}, {Errani}, \& {Read}}]{2025arXiv251215886V}
{Valenciano}, F., {Camalich}, J.~M., {Di Cintio}, A., {et~al.} 2025, arXiv
  e-prints, arXiv:2512.15886, \dodoi{10.48550/arXiv.2512.15886}

\bibitem[{{Vasiliev}(2019)}]{2019MNRAS.482.1525V}
{Vasiliev}, E. 2019, \mnras, 482, 1525, \dodoi{10.1093/mnras/sty2673}

\bibitem[{{Virtanen} {et~al.}(2020){Virtanen}, {Gommers}, {Oliphant},
  {Haberland}, {Reddy}, {Cournapeau}, {Burovski}, {Peterson}, {Weckesser},
  {Bright}, {van der Walt}, {Brett}, {Wilson}, {Millman}, {Mayorov}, {Nelson},
  {Jones}, {Kern}, {Larson}, {Carey}, {Polat}, {Feng}, {Moore}, {VanderPlas},
  {Laxalde}, {Perktold}, {Cimrman}, {Henriksen}, {Quintero}, {Harris},
  {Archibald}, {Ribeiro}, {Pedregosa}, {van Mulbregt}, \& {SciPy 1. 0
  Contributors}}]{2020NatMe..17..261V}
{Virtanen}, P., {Gommers}, R., {Oliphant}, T.~E., {et~al.} 2020, Nature
  Medicine, 17, 261, \dodoi{10.1038/s41592-019-0686-2}

\end{thebibliography}



\end{document}